\documentclass[aps,prb,reprint,showpacs,amsmath,amssymb,superscriptaddress,citeautoscript]{revtex4-1}
\usepackage{graphicx}% Include figure files
\begin{document}

%\preprint{E. Lengyel et al.}

\title{Avoided ferromagnetic quantum critical point in CeRuPO}

\author{E. Lengyel} \email{lengyel@cpfs.mpg.de}\affiliation{Max Planck Institute for Chemical Physics of Solids, N\"{o}thnitzer Str.\ 40, 01187 Dresden, Germany}

\author{M. E. Macovei} \affiliation{Max Planck Institute for Chemical Physics of Solids, N\"{o}thnitzer Str.\ 40, 01187 Dresden, Germany}

\author{A. Jesche} \affiliation{EP VI, Center for Electronic Correlations and Magnetism, Augsburg University, 86159 Augsburg, Germany}

\author{C. Krellner} \affiliation{Institute of Physics, Goethe University Frankfurt, Max-von-Laue-Strasse 1, 60438
Frankfurt am Main, Germany}

\author{C. Geibel} \affiliation{Max Planck Institute for Chemical Physics of Solids, N\"{o}thnitzer Str.\ 40, 01187 Dresden, Germany}

\author{M. Nicklas} \affiliation{Max Planck Institute for Chemical Physics of Solids, N\"{o}thnitzer Str.\ 40, 01187 Dresden, Germany}

\date{\today}
\begin{abstract}

CeRuPO is a rare example of a ferromagnetic (FM) Kondo-lattice system. External pressure suppresses the ordering temperature to zero at about $p_c\approx3$ GPa. Our ac-susceptibility and electrical-resistivity investigations evidence that the type of magnetic ordering changes from FM to antiferromagnetic (AFM) at about $p^*\approx0.87$\,GPa. Studies in applied magnetic fields suggest that ferromagnetic and antiferromagnetic correlations compete for the ground state at $p>p^*$, but finally the AFM correlations win. The change in the magnetic ground-state properties is closely related to the pressure evolution of the crystalline-electric-field level (CEF) scheme and the magnetic Ruderman-Kittel-Kasuya-Yosida (RKKY) exchange interaction. The N\'{e}el temperature disappears abruptly in a first-order-like fashion at $p_c$, hinting at the absence of a quantum critical point. This is consistent with the low-temperature transport properties exhibiting Landau-Fermi-liquid (LFL) behavior in the whole investigated pressure range up to 7.5\,GPa.

\end{abstract}

\pacs{71.27.+a,75.30.Kz,74.70.Xa}

\maketitle

\section{INTRODUCTION}

The delicate interplay between competing interactions in strongly correlated metals might lead to the emergence of novel unconventional phases which defy the standard theories of metals. In Ce or Yb $4f$-based Kondo-lattice systems the Kondo effect, leading to the formation of a non-magnetic Kondo singlet, competes with the magnetic Ruderman-Kittel-Kasuya-Yosida (RKKY) interaction. External or chemical pressure offer an immediate way to tune the interplay of these energy scales. A particularly successful way to find new unconventional states in these materials is to continuously suppress a magnetically ordered state to zero temperature at a quantum critical point (QCP). In the vicinity of this QCP, often unconventional metallic phases emerge,which display deviations from Landau-Fermi-liquid (LFL) behavior or unconventional superconducting states emerge. While many antiferromagnetic (AFM) Kondo-lattice systems are known, only a few show a ferromagnetic (FM) ground state, e.g.\ YbNi$_4$P$_2$,\cite{Krellner11,Steppke13} CePd$_{1-x}$Rh$_x$,\cite{Sereni07} CeAgSb$_{2}$\cite{sidorov03}, or CeRuPO\cite{krellner07,krellner08}, and are, therefore, of special interest.

CeRuPO is a Kondo-lattice system ($T_K\sim10$~K) which orders ferromagnetically at $T_C=15$~K \cite{krellner07,krellner08}. It belongs to the Ce$T$PO ($T$ = Fe, Ru, Co, and Os) family of compounds, which crystallizes in the tetragonal ZrCuSiAs-type structure.\cite{zimmer95} The Ce$T$PO compounds share the same crystal structure as the '1111' high-temperature iron-pnictide superconductors. The connection between the two families has been exemplified in CeFePO. By substituting phosphorous with arsenic, one obtains CeFeAsO, a parent compound of the iron-pnictide superconductors.\cite{Jesche12,Mydeen12} Ce$T$PO can be seen as consisting of alternating layers of $T$P$_4$ and OCe$_4$ tetrahedra stacked along the crystallographic $c$ direction. This layered structure is also reflected in the quasi-two dimensional character of the Fermi surface found in band-structure calculations.\cite{krellner07} The magnetization  evidences a strongly anisotropic behavior with respect to the magnetic-field direction in the FM state. For field applied perpendicular to the $c$ axis a saturation moment of $\mu^{ab}_{\rm sat}=1.2\,\mu_{\rm B}$, and no hysteresis was found.\cite{krellner08} In contrast, for field applied parallel to the $c$ axis, a spontaneous magnetic moment of $0.3\,\mu_{\rm B}$ was observed. The saturation moment is only $\mu^{c}_{\rm sat}=0.43\,\mu_{\rm B}$. The hysteresis loop with a coercive field of $\mu_{0}H_{c}=0.1$\,T confirms the FM order. The magnetization data indicate that the magnetic moments are aligned ferromagnetically along the $c$ direction, while the easy magnetization direction is perpendicular to it. This behavior can be interpreted as a competition between the crystalline-electric-field (CEF) anisotropy and the anisotropy of the RKKY exchange interaction with respect to the $x$, $y$, and $z$ components of the magnetic moment \cite{krellner08}. The RKKY interaction, which is responsible for the FM ordering is larger for the $z$ component of the magnetic moment than for the $x$ and $y$ components, while the CEF anisotropy leads to a ground state CEF doublet with a larger saturation moment in the basal plane than along the crystallographic $c$ axis. This complex magnetic behavior at ambient pressure suggests a strong sensitivity of the magnetic ground state to external parameters, such as chemical or hydrostatic pressure. Indeed, isoelectronic substitution of Ru by Fe in CeRu$_{1-x}$Fe$_x$PO suppresses the FM state continuously to zero temperature at about $x\approx0.86$, which suggests the existence of a FM QCP.\cite{Kitagawa12,Kitagawa13,Lausberg12}

Previous electrical-resistivity studies under external pressure suggest that the magnetic ordering temperature is suppressed to zero around 3~GPa.\cite{Macovei09,Kotegawa13} Based on resistivity data in a static magnetic field, it was speculated that the type of the magnetic ordering changes from ferromagnetic to antiferromagnetic as a function of pressure.\cite{Kotegawa13} Here, we present a combined study of magnetic and electrical-transport properties evidencing the change of the type of magnetic ordering from FM to AFM at $p^*\approx0.87$\,GPa. The analysis of our data substantiates the competition between FM and AFM correlations. We further find that the N\'{e}el temperature disappears abruptly around 3~GPa. This observation is consistent with the absence of any signature of the existence of a QCP in the temperature dependence of the resistivity.

This paper is organized as follows. In Sec.\,\ref{Methods}, we describe the sample preparation and characterization as well as the details of the electrical-resistivity and ac-susceptibility experiments under pressure. We present our experimental results in Sec.\,\ref{results}. After describing the magnetic-susceptibility data obtained at various pressures in Sec.\,\ref{susceptibility}, we establish the temperature -- pressure phase diagram of CeRuPO in Sec.\,\ref{phase diagram}. In Sec.\,\ref{magnetic field} we study the effect of the application of a static magnetic field on the different types of ground states found in CeRuPO under pressure. Finally, we analyze the low-temperature electrical resistivity in Sec.\,\ref{resistivity}. The discussion of our results in Sec.\,\ref{discussion} is followed by the summary in Sec.\,\ref{summary}. Appendix \ref{apx_mag} describes the details of the analysis of the low-temperature resistivity in the ordered states and Appendix\ \ref{spinwave} provides additional information on the spin-wave gap and stiffness in the AFM state. The effect of a magnetic field on the low-temperature properties in the paramagnetic (PM) state is presented in the Appppendix\ \ref{parmagnetic}.

\section{METHODS}\label{Methods}

 CeRuPO single crystals were prepared using a Sn-flux technique \cite{krellner08}. The crystals have the form of platelets with metallic luster. Electron-microprobe and energy-dispersive x-ray analysis revealed the formation of single-phase CeRuPO single crystals with a stoichiometric Ce:Ru:P content and the presence of the oxygen. X-ray powder diffraction patterns confirm the ZrCuSiAs structure type with lattice parameters $a=4.027(3)$\,{\AA} and $c=8.256(2)$\,{\AA} \cite{krellner08}, in good agreement with those reported in the literature \cite{zimmer95,krellner07}.

 A double-layered piston-cylinder type pressure cell has been used for generating pressures up to 3~GPa. In order to achieve hydrostatic pressure conditions, silicon-oil served as the pressure-transmitting medium. This type of pressure cell was used for both, electrical-resistivity, $\rho$, and ac-susceptibility, $\chi_{ac}$, measurements. For electrical-resistivity experiments at pressures up to $10$\,GPa a Bridgman-type pressure cell using steatite as the pressure-transmitting medium was utilized. The pressure was monitored at low temperatures by the shift of the superconducting transition temperature of a piece of Pb placed inside the pressure chamber \cite{Eiling81}. From the width of the transition, a pressure gradient smaller than 0.027~GPa could be estimated for the piston-cylinder type pressure cell in the entire pressure range, while in the Bridgman-type pressure cell, a gradient of up to 10\% was observed. Temperatures down to 1.8~K and 50~mK could be achieved in a Physical Properties Measurement System (Quantum Design) and a dilution refrigerator (Oxford Instruments), respectively. Both systems were equipped with superconducting magnets generating fields up to 9~T. A Linear Research LR700 resistance/mutual-inductance bridge working at a measuring frequency of 16~Hz was utilized for the ac-susceptibility and the four-probe resistivity measurements. Additionally, a Lock-in amplifier together with a low-temperature transformer were employed for the resistivity experiments in the dilution refrigerator, for which a frequency of 113~Hz for the electrical current was used. The $\chi_{ac}$ data at different pressures were normalized using the jump height of the SC phase transition of the piece of Pb placed next to the sample, which was primarily used as a pressure sensor. In this way, we can compare the size of the susceptibility signal taken at different pressures, despite the arbitrary units of the real and the imaginary parts of the susceptibility.

\section{EXPERIMENTAL RESULTS}\label{results}

\subsection{Magnetic susceptibility}\label{susceptibility}

\begin{figure}[b!]
    \centering
    \includegraphics [width=0.95\linewidth]{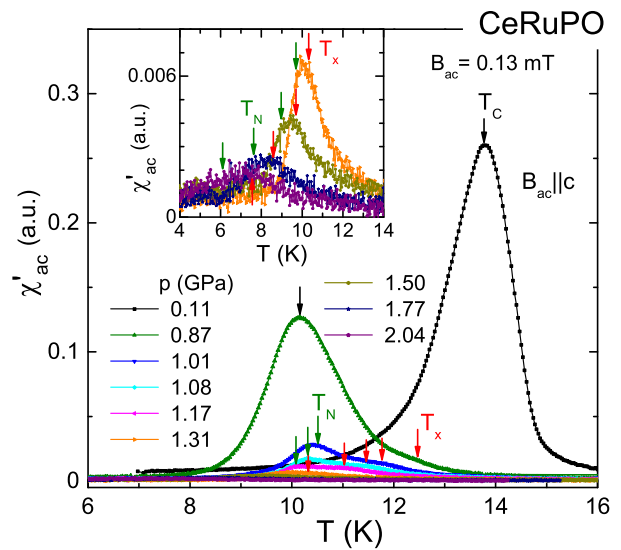}
    \renewcommand{\baselinestretch}{1.2}
    \caption{Temperature dependence of the ac susceptibility, $\chi{'}_{ac}$, of CeRuPO at various pressures. The transition temperatures, $T_{C}$ and $T_{N}$, are indicated by arrows. The origin of the anomaly at $T_{X}$ is not clear (see text for details). For clarity, the $\chi{'}_{ac}(T)$ curves for $1.31\,\rm GPa \leq p \leq 2.04\,\rm GPa$ are enlarged in the inset.}
    \label{chi(p)}
\end{figure}

\begin{figure}[t!]
    \centering
    \includegraphics [width=0.95\linewidth]{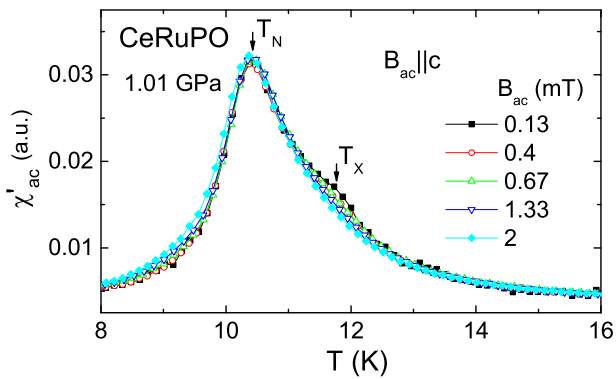}
    \renewcommand{\baselinestretch}{1.2}
    \caption{Temperature dependence of the real component of the ac susceptibility of CeRuPO at $p=1.01\,\rm GPa$ as a function of the ac field. The arrows mark the AFM ordering temperature $T_{N}$ and the anomaly at $T_{X}$.}
    \label{Ce101GPachivsBac}
\end{figure}

The results of the ac-susceptibility measurements on CeRuPO for selected pressures are shown in Fig.\,\ref{chi(p)}. The ac field was applied parallel to the crystallographic $c$ axis, in the direction of the ordered magnetic moments \cite{krellner08}. At $p=0.11$\,GPa, a rapid increase in $\chi{'}_{ac}(T)$ with lowering the temperature signals the transition from a PM to a FM state. The maximum of the sharp peak is taken to define the ordering temperature. The increase of pressure leads to the suppression of the ordering temperature. At $p=1.01$\,GPa the amplitude of the $\chi{'}_{ac}(T)$ peak is strongly reduced when compared to lower pressures, suggesting that the type of magnetic order is changed from FM to AFM. At this pressure, $T_{N}=10.44$\,K is obtained. For pressures up to $2.04$\,GPa, a hump in $\chi{'}_{ac}(T)$ above the ordering temperature points to the existence of a second anomaly. We denote the temperature characteristic to this anomaly as $T_{X}$. The presence of this anomaly is observed in both, real and imaginary parts of the ac susceptibility. Furthermore, this anomaly shifts to lower temperatures as pressure increases and is suppressed by applying a small static magnetic field of $B=0.01$\,T or by increasing the modulation field, $B_{ac}$, (see Fig.\,\ref{Ce101GPachivsBac} and the corresponding discussion below). Currently, the origin and the nature of the anomaly at $T_{X}$ are not known, but the magnetic-field dependence of the ac susceptibility described above might be explained by the formation of short-range FM correlations at $T_{X}$. As expected for Ce-based systems, the magnetic order is suppressed by pressure, thus at $2.04$\,GPa the magnetic order is detected in $\chi{'}_{ac}(T)$ at $T_{N}=6.1$\,K. This is the highest pressure where we unambiguously detect the phase transition in our ac-susceptibility measurements, and thus the values of $T_N$ obtained from ac-susceptibility data above this pressure have to be considered with relatively large error bars. However, we can clearly follow the magnetic transition to pressures up to nearly $3$\,GPa in our electrical-resistivity measurements, where the anomaly is very sharp. Similar to $\chi{'}_{ac}(T)$, the imaginary component of the susceptibility shows a significant change by application of pressure. For $p\geq1.01$\,GPa, no anomaly corresponding to the magnetic transition can be observed in $\chi{''}_{ac}(T)$, in contrast to the very large peak which we detect at lower pressures. This supports our suggestion of a pressure-induced change from FM to AFM order in CeRuPO.

\begin{figure}[t!]
    \centering
    \includegraphics[width=0.95\linewidth,clip]{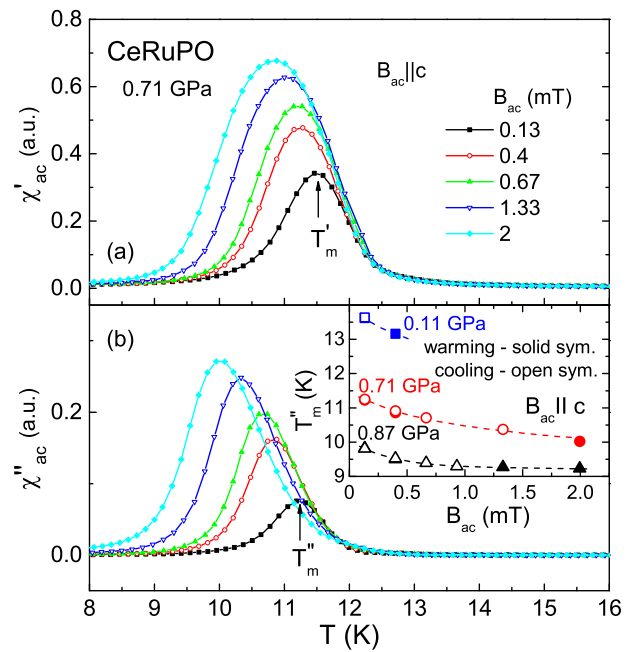}
    \renewcommand{\baselinestretch}{1.2}
   \caption{Temperature dependence of the (a) real and (b) imaginary parts of the ac susceptibility of CeRuPO at $p=0.71\,\rm GPa$ at different modulation fields $B_{ac}$. The evolution of the position of the peak in $\chi{''}_{ac}(T)$ with $B_{ac}$ is displayed in the inset of (b) for various pressures.}
   \label{Ce071GPachivsBac}
\end{figure}

Additional information on the nature of the magnetic transition is obtained from changing the driving field (see, for example, Ref.\,\onlinecite{Levin01}). Since a FM material below $T_{C}$ responds to magnetic field changes by domain-wall movement (DWM), an increase in the amplitude of the ac-modulation field leads to a change of the position of the maximum and of the width of the anomaly in $\chi_{ac}(T)$. At an AFM transition, this effect is not present. In CeRuPO at $p=0.71\,\rm GPa$, the influence of the ac field on both the real and imaginary components of the ac susceptibility is shown in Figs.\,\ref{Ce071GPachivsBac}(a) and \ref{Ce071GPachivsBac}(b), respectively. In the PM state, all magnetic-susceptibility curves are indistinguishable, whereas in the magnetic state, the temperature dependencies of both $\chi{'}_{ac}$ and $\chi{''}_{ac}$ display nearly symmetric peaks with maxima at $T^{'}_{m}$ and $T^{''}_{m}$, respectively. Upon increasing the ac-field amplitude from $0.13$ to $2$\,mT, the magnitude of the $\chi{'}_{ac}$ peak increases by a factor of $2$, whereas that of the $\chi{''}_{ac}$ peak increases by a factor of $3.5$. The positions of both peaks shift toward lower temperatures as the ac field increases. However, the dependence of $T^{''}_{m}$ on $B_{ac}$ in the case of $\chi{''}_{ac}$ is stronger as compared to $\chi{'}_{ac}$ ($T^{'}_{m}$). The $B_{ac}$ dependence of $T^{''}_{m}$ for selected pressures where the system has a FM ground state is presented in the inset of Fig.\,\ref{Ce071GPachivsBac}(b). The imaginary component of the ac susceptibility reflects the energy-loss processes in a FM material due to energy absorption by the domain walls during their excitation. Thus, the shift of the $\chi{''}_{ac}$ peak in CeRuPO to lower temperatures with increasing the ac field indicates that at, e.g., $p=0.71\,\rm GPa$, $B_{ac}=2$\,mT is strong enough to excite the DWM at $10$\,K, while an ac field of $0.13$\,mT can only excite DWM at higher temperatures of about $\sim11.25$\,K. Furthermore, the nearly exponential dependence of $T^{''}_{m}$ versus the amplitude of the ac field [inset of Fig.\,\ref{Ce071GPachivsBac}(b)] indicates that the excitation of the domain walls is thermally activated. All of the properties are similar to those reported in standard FM systems and confirm the FM nature of the ordering. The $B_{ac}$ dependence of $T^{''}_{m}$ is getting less and less pronounced upon increasing pressure. This suggests that DWM in CeRuPO is hindered by the increase of pressure.

In contrast to the situation at $p=0.71$\,GPa, Fig.\,\ref{Ce101GPachivsBac} depicts the temperature dependence of the real part of the ac susceptibility of CeRuPO at $p=1.01$\,GPa at different ac fields. Here, the $\chi{'}_{ac}$ peak at $T_{N}$ is independent of the driving field, as expected for an AFM phase transition. The weak shoulder above $T_N$ is attributed to the anomaly at $T_X$. As mentioned earlier, the anomaly observed at $T_{X}$ is suppressed by increasing $B_{ac}$. It is worth mentioning that for $p\geq1.01$\,GPa, the peak size in $\chi{''}_{ac}(T)$ (not shown) is drastically reduced when compared to lower pressures and only the position of the peak has a very weak ac-field dependence. Based on the temperature where this peak appears, we attribute this weak signal in $\chi{''}_{ac}(T)$ to the anomaly at $T_{X}$, concluding that no signature of the magnetic phase transition in the imaginary part of the ac susceptibility exists when the system turns from the PM to the AFM ground state.

\subsection{Temperature -- pressure phase diagram}\label{phase diagram}

\begin{figure}[t!]
    \centering
    \includegraphics [width=.95\linewidth]{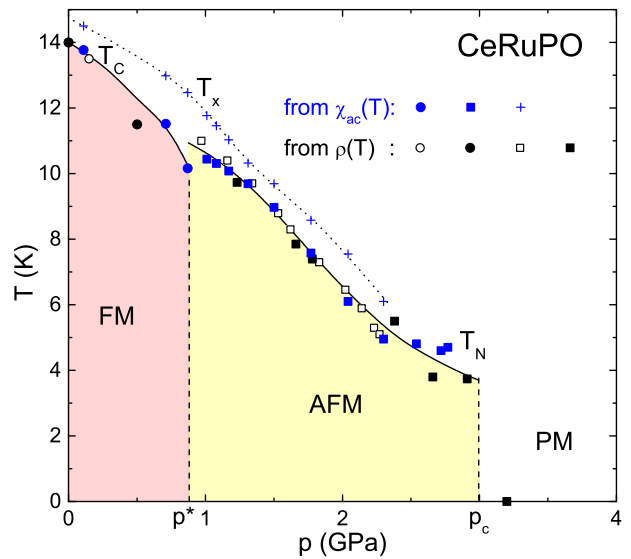}
    \renewcommand{\baselinestretch}{1.2}
    \caption{Temperature -- pressure phase diagram of CeRuPO. In addition to the results from our ac-susceptibility measurements, data from resistivity experiments in a piston-cylinder-type cell (open symbols, Ref.~\onlinecite{Macovei09}) and a Bridgman-type pressure cell (closed symbols, this work) have been included. The magnetic-ordering temperatures, $T_{C}$ and $T_{N}$, are indicated by circles and squares, respectively. The anomaly observed in $\chi_{ac}$ measurements at $T_{X}$ is indicated by $+$. The dashed lines at $p^{*}$ and $p_{c}$ mark a change in the ground state of the system from FM to AFM and AFM to PM, respectively. The lines are guides to the eye.}
    \label{PhD}
\end{figure}

The pressure dependencies of the magnetic transitions in CeRuPO are depicted in Fig.\,\ref{PhD}. We included in this figure our data from both electrical-resistivity and magnetic-susceptibility measurements. As one can see, there is a good agreement between the results obtained from $\rho(T)$ measurements using piston-cylinder-type (open symbols, Ref.\,\onlinecite{Macovei09}) and Bridgman-type (closed symbols) pressure cells and those from the $\chi_{ac}(T)$ experiments. The FM order observed at ambient pressure at $14$\,K shifts to lower temperatures upon applying pressure, so that at $p=0.87$\,GPa the system orders ferromagnetically at $T_{C}=10.16$\,K. With further increasing pressure ($p\geq1.01$\,GPa), the nature of the magnetic order changes to AFM. Thus, we estimate the pressure at which this change occurs to be $p^{*} \approx 0.87$\,GPa. The N\'{e}el temperature is suppressed continuously by increasing pressure to $T_{N}= 4.3$\,K at $p=2.9$\,GPa. Above this pressure, i.e., at $p=3.2$\,GPa, no indication of any magnetic transition can be observed in $\rho(T)$ down to the lowest accessible temperature of $50$\,mK, suggesting that the magnetic order is suppressed in a first-order-like fashion. Therefore, we estimated the critical pressure to be $p_{c} \approx 3$\,GPa. The similar slopes observed in the pressure dependencies of the ordering temperatures in the FM and AFM states hint at the same energy scale determining the size of the magnetic-transition temperature.

\subsection{Effect of magnetic field}
\label{magnetic field}

We have also investigated the effect of a static magnetic field on the ground-state properties of CeRuPO. In the following, we primarily focus on the magnetic-field dependence of the ac susceptibility. However, our conclusions are also supported by our electrical-resistivity measurements. The three distinct regions emphasized in Fig.\,\ref{PhD} by FM ($p<p^{*}$), AFM ($p^{*}<p< p_{c}$), and PM ($p>p_{c}$) will be discussed separately (see Appendix\,\ref{parmagnetic} for the PM region).

\subsubsection{Ferromagnetic state}

\begin{figure}[t!]
  \centering
 \includegraphics[width=.95\linewidth]{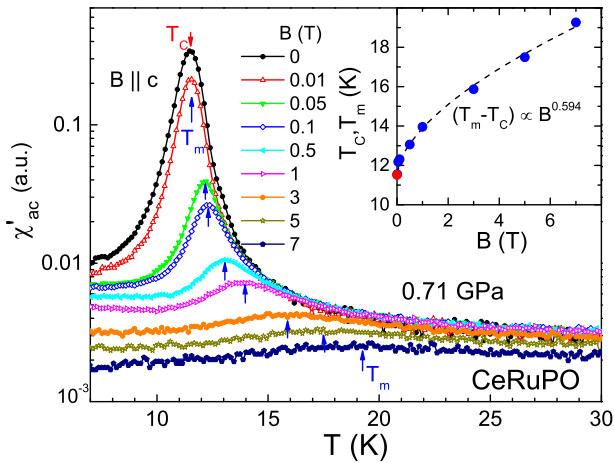}
  \renewcommand{\baselinestretch}{1.2}
  \caption{Temperature dependence of the real component of the susceptibility, $\chi{'}_{ac}$, of CeRuPO at $p= 0.71\,\rm GPa$ at different applied magnetic fields ($B\parallel c$). At $B=0$, the ordering temperature $T_{C}$ is marked by an arrow. For $B>0$, $\chi{'}_{ac}(T)$ passes through a maximum which is denoted by $T_{m}$. Inset: $T_{m}$ {\it vs.} magnetic field.}
 \label{Ce071GPa_re_vs_B}
\end{figure}

We start with the results in the FM region. Figure\,\ref{Ce071GPa_re_vs_B} shows the temperature dependence of the real component of the susceptibility, $\chi{'}_{ac}$, at $p=0.71\,\rm GPa$ at different static magnetic fields ($B\parallel c$). At $B=0$ and $0.01$\,T, the susceptibility depends on the amplitude of the ac\ field, while for $B\geq0.05$\,T, such a behavior is no longer observed. Therefore, in Fig.\,\ref{Ce071GPa_re_vs_B} the curves at $B=0$ and $0.01$\,T were measured with $B_{ac}=0.13$\,mT, while for $B\geq0.05$\,T, $B_{ac}$ was increased to $0.4$\,mT, allowing a better resolution of the measured signal. At $B=0$, $\chi{'}_{ac}(T)$ increases rapidly with decreasing temperature as the FM transition at $T_{C}= 11.52$\,K is approached, and then decreases as the temperature is lowered further. For $B>0$, $\chi{'}_{ac}(T)$ displays a peak structure similar to that observed at $B=0$. $T_{m}$ is assigned to the position of the maximum of these curves. As the applied static field increases, the peak decreases in amplitude, becomes broader, and moves to higher temperatures. This is exactly the behavior expected for a ferromagnet where application of a longitudinal field results in a change from a second-order phase transition at $B=0$ to a crossover at finite field.
It is similar with the findings observed at $p=0$ \cite{krellner08}. The evolution of $T_{m}$ with applied magnetic field shows a $(T_{m}-T_{C})\propto B^{\lambda}$ dependence (inset of Fig.\,\ref{Ce071GPa_re_vs_B}). A fit of the $T_{m}(B)$ data yields $\lambda=0.594$. The imaginary component of the susceptibility, $\chi{''}_{ac}$ (not shown), has a strong $T$ dependence for small static magnetic fields ($B\leq 0.01$\,T), reflecting the energy losses caused by domain-wall movements, whereas an almost zero value of $\chi{''}_{ac}$ is detected at $B\gtrsim 0.05$\,T, reflecting the reversible magnetization increment for a field larger than the coercive field.

\subsubsection{Antiferromagnetic state}\label{AFM_state}

\begin{figure}[t!]
\centering
\includegraphics[width=0.95\linewidth]{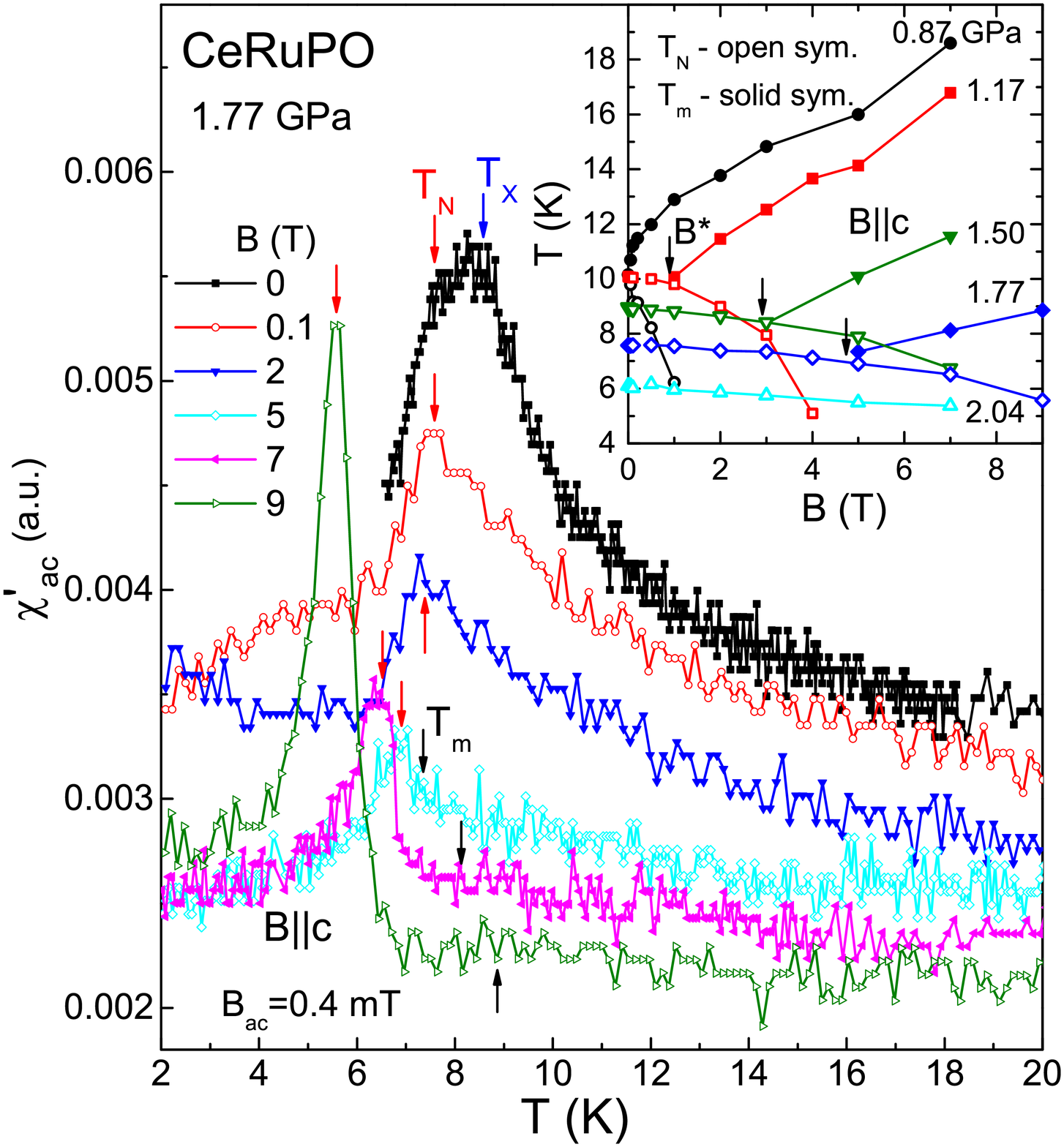}
\renewcommand{\baselinestretch}{1.2}
\caption{Temperature dependence of the real part of the magnetic susceptibility of CeRuPO at $p=1.77\,\rm GPa$ at different magnetic fields ($B\parallel c$) in the antiferromagnetic region of the $T-p$ phase diagram. The arrows mark the AFM phase transition at $T_{N}$, and the crossover temperatures $T_{m}$ and $T_{X}$. Inset: $T-B$ phase diagram for different pressures, all (except 0.87\,GPa) in the antiferromagnetic regime.}
\label{Ce177G_re_vs_B}
\end{figure}

In the following, we move to the AFM region of the $T-p$ phase diagram of CeRuPO (Fig.\,\ref{PhD}). The temperature dependence of the real component of the magnetic susceptibility at $p=1.77$\,GPa in various static magnetic fields applied parallel to the crystallographic $c$ axis is depicted in Fig.\,\ref{Ce177G_re_vs_B}. At $B=0$, the AFM transition is seen in $\chi{'}_{ac}(T)$ as a peak with the maximum at $T_{N}=7.58$\,K. For low magnetic fields, the peak decreases in amplitude and moves toward lower temperatures as the applied magnetic field increases.
By comparing the $\chi{'}_{ac}$ data at higher fields with corresponding data obtained at lower pressures, $p \leq p^*$, (see Fig.\,\ref{Ce071GPa_re_vs_B}), one can identify a very broad maximum at $T_m\approx 8.6$~K at 9~T, which shifts toward lower temperatures with decreasing the magnetic field and gets absorbed in the $T_N$ anomaly for magnetic fields slightly below 5~T. This behavior is very similar to that observed at fixed magnetic field as a function of pressure, which will be discussed in Sec.\,\ref{PhD_field} [see also Fig.\,\ref{all_1T}(a)] and thus hints at the presence of FM correlations. Starting from the value of the magnetic field where the anomaly at $T_{m}$ is first observed, $B^*$, the peak in $\chi{'}_{ac}(T)$ related to the AFM transition sharpens, suggesting a first-order-like transition from a field-polarized FM to an AFM order.

The changes of $T_{N}$ and $T_{m}$ with magnetic field for selected pressures between $p=0.87$\,GPa and $p=2.04$\,GPa are illustrated in the $T-B$ phase diagram shown in the inset of Fig.\,\ref{Ce177G_re_vs_B}. At $p=0.87$\,GPa, close to the pressure, $p^*$, we detect an AFM phase slightly below $T_C$, which is suppressed very rapidly by a magnetic field and replaced by a phase with FM correlations. The crossover line $T_{m}$ which separates this nearly FM phase from the PM phase shifts to higher temperatures with increasing field, and thus gets stabilized by the external field.
With increasing pressure, the AFM regime extends to higher magnetic fields, while the nearly FM phase gets suppressed, with respect to both the PM and the AFM phase. The nearly FM phase is observed only above a critical field $B^*$, which increases steadily with pressure to above our maximum field at $p=2.04$\,GPa. The slope of $T_m(B)$, which reflects the size of the moment, decreases only weakly with increasing pressure, indicating a slight decrease of the size of the moment with increasing pressure. In contrast, the absolute value of the slope of $T_N(B)$ strongly decreases with increasing pressure, indicating a strong increase of the AFM coupling with respect to the FM one.

\begin{figure}[t!]
\centering
\includegraphics[width=0.95\linewidth]{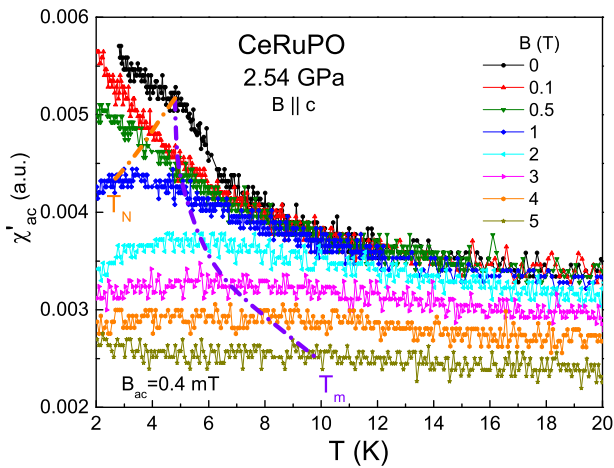}
\renewcommand{\baselinestretch}{1.2}
\caption{Temperature dependence of the real part of the magnetic susceptibility of CeRuPO at $p=2.54\,\rm GPa$ at different magnetic fields, $B\parallel c$. Due to the low accuracy of the data we only trace the magnetic-field dependencies of the anomalies at $T_{N}$ and $T_{m}$ by the dashed-dotted lines.}
\label{Ce254G_re_vs_B}
\end{figure}

The situation changes at pressures above 2.04\,GPa. There, the accuracy in identifying the magnetic-ordering temperatures from $\chi{'}_{ac}(T)$ measurements is considerably reduced. However, the electrical-resistivity data support our results. Figure\,\ref{Ce254G_re_vs_B} shows as an example $\chi{'}_{ac}(T)$ of CeRuPO at $p = 2.54$\,GPa at various static magnetic fields, $B \parallel c$. At $B = 0$, a phase transition into the antiferromagnetically ordered state is visible at $T_N = 4.81$\,K. Application of a static magnetic field suppresses $T_N$ and a broad and weak maximum forms in $\chi{'}_{ac}(T)$ at larger magnetic fields. The maximum washes out and moves toward higher temperatures with increasing the magnetic field. This high magnetic field maximum at $T_m$ is about $3 - 5$ times less pronounced than that observed for the same magnetic field at $p \leq 2.04$\,GPa and it is identifiable already around $1-2$\,T, whereas that at lower pressures appeared at $B > B^*$, with $B^*$ increasing upon increasing the pressure and $B^* > 7$\,T at $p = 2.04$\,GPa. Due to a large uncertainty in determining the transition/crossover temperatures, we only suggest the magnetic-field evolution of these anomalies by the dash-dotted lines in Fig.\,\ref{Ce254G_re_vs_B}. However, even if $T_N$ and $T_m$ cannot be unambiguously resolved, the increase of $T_m$ upon increasing $B$ is obvious in the data and might be related with the presence of a ferromagnetically polarized phase. We want to mention that this high-field anomaly at $p > 2.04$\,GPa is visible in both, our electrical-resistivity, where $B \perp c$, and ac-susceptibility ($B\parallel c$) measurements.

Notably, despite an AFM ground state, a negative temperature dependence is observed in $\chi{'}_{ac}(T)$ at low temperatures (below $\approx 5$\,K) at low magnetic fields, $B \leq 0.5$\,T, for $p\geq2.3$\,GPa. This negative temperature dependence of $\chi{'}_{ac}$ hints at the presence of FM-like correlations. Altogether, these results suggest the existence of FM correlations in CeRuPO in the pressure range where the AFM order becomes weaker, $p > 2.04$\,GPa. These correlations seem to be different in nature from those existing at lower pressures. Thus, we suggest that a change in the magnetic ground state of CeRuPO takes place just above $2.04$\,GPa. As observed in several FM Kondo-lattice systems, e.g., CeFePO,\cite{Lausberg12} CePd$_{1-x}$Rh$_x$,\cite{Westerkamp08} and CeTiGe$_3$ \cite{Kittler13}, one might consider FM short-range or spin-glass-like order in this pressure range.

Since the magnetic anisotropy plays an important role in CeRuPO \cite{krellner08}, we have also investigated the pressure evolution of the anisotropy in this material. Our pressure dependent electrical-resistivity measurements performed with $B\parallel c$ and $B\perp c$ (not shown) at 1.8\,GPa evidence that the evolutions of $T_{N}$ and $T_{m}$ with magnetic field for both directions of the magnetic field are similar. Within the limits of the error bars, there is no difference in the $T_{N}(B)$ and $T_{m}(B)$ curves for field applied along the different crystallographic directions. In contrast to this, at $p=0$, the anomaly in the resistivity associated with the magnetic transition, $T_{m}$, shows a stronger field dependence for magnetic field applied perpendicular to the crystallographic $c$ axis than for $B\parallel c$ \cite{krellner08}. These results suggest that the anisotropy in CeRuPO is reduced at higher pressures. The good agreement between the results obtained from resistivity with $B \perp c$ and from ac susceptibility where $B \parallel c$ for $p^* < p < 3$\,GPa also supports this conclusion. We speculate that the decrease of the anisotropy under pressure might cause the change from FM to AFM order at $p^*$.

\subsubsection{Evolution of the $T-p$ phase diagram in magnetic field} \label{PhD_field}

\begin{figure}[t!]
\centering
\includegraphics[width=0.95\linewidth]{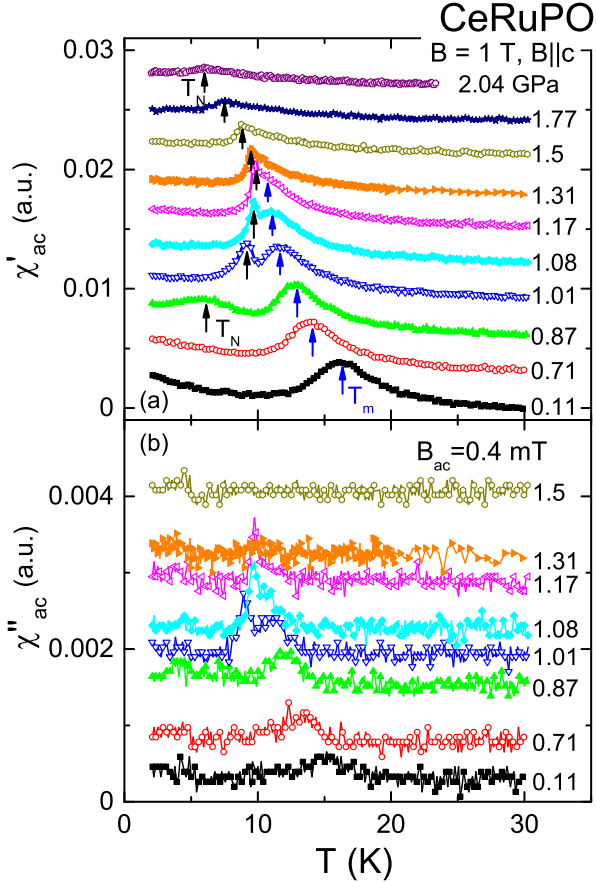}
\renewcommand{\baselinestretch}{1.2}
\caption{Temperature dependence of the (a) real and (b) imaginary parts of $\chi_{ac}(T)$ of CeRuPO in $B=1$\,T ($B\parallel c$) at different $p$. The arrows mark the ordering/crossover temperatures as indicated. For clarity, in both figures the data are shifted on the vertical axis.}
\label{all_1T}
\end{figure}

In order to get a better view of the $T-p-B$ phase diagram of CeRuPO, we compare in Fig.\,\ref{all_1T} the temperature dependencies of the real and imaginary parts of the ac\ susceptibility at different pressures for a fixed field of $B = 1$\,T. For $p\leq0.71$\,GPa, only the anomaly at $T_{m}$ corresponding to the crossover from a PM to a ferromagnetically polarized state is detected in both $\chi{'}_{ac}$ and $\chi{''}_{ac}$. For $p^{*}\leq p\lesssim 1.3$\,GPa, a second anomaly can be observed below $T_{m}$, denoted as  $T_{N}$. This reflects the transition from a field-polarized FM to an AFM order. In this pressure range, $T_{m}(p)$ continues to decrease, while $T_{N}(p)$ increases with increasing pressure. As $T_{N}(p)$ and $T_{m}(p)$ are coming closer to each other with increasing the pressure, the anomaly in $\chi{'}_{ac}$ at $T_{N}$ is sharpening, suggesting that the transition from the field-polarized FM to the AFM state becomes first order. This is also reflected in the imaginary part of the susceptibility [see Fig.\,\ref{all_1T}(b)]. Here, the anomaly at $T_{N}$ cannot be observed at $p \approx 0.87$\,GPa, but it becomes evident for $p \geq 1.01$\,GPa, supporting a first-order nature of the transition from the field-polarized FM to the AFM order. Above $p \approx 1.3$\,GPa, only the phase transition at $T_{N}$ can be detected in the real part of the susceptibility at $B=1$\,T. The shape of the anomaly at the phase transition in $\chi{'}_{ac}(T)$ and the fact that no corresponding anomaly can be resolved in $\chi{''}_{ac}(T)$, are hinting at a second-order AFM phase transition in this pressure range.

\begin{figure}[t!]
\centering
\includegraphics[width=0.95\linewidth]{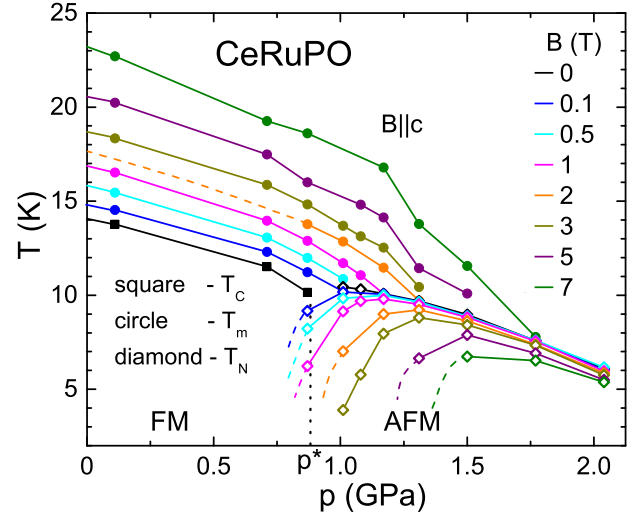}
\renewcommand{\baselinestretch}{1.2}
\caption{Magnetic $T-p$ phase diagram of CeRuPO compiled from ac-susceptibility data at different magnetic fields. The FM and AFM ground states at $B=0$ are indicated. $p^*$ marks the pressure where the type of magnetic ordering changes.}
\label{T-p_field}
\end{figure}

By summarizing our results for various static magnetic fields up to $B = 7$\,T, we build the $T-p$ phase diagram for different values of $B$. The results are presented in Fig.\,\ref{T-p_field}. Due to the low accuracy in determining the temperatures of the anomalies observed at $p > 2.04$\,GPa, we limit ourselves to the data obtained for $p \leq 2.04$\,GPa. Figure~\ref{T-p_field} clearly shows the transition from a FM to an AFM ground state at $p^{*}\approx 0.87$\,GPa and the stabilization of the FM state in an applied magnetic field. Guided by the pressure dependence of the FM crossover lines, $T_M(p,B)$, in various magnetic fields, we extrapolate $T_C(p)$ at zero field to a QCP at $T=0$ at $p_{_{\rm FM}}\approx 1.6$\,GPa hidden by the AFM phase. This pressure increases gradually with increasing the magnetic field. At $B = 7$\,T, we find $p_{_{\rm FM}}\approx 2$\,GPa.

\subsection{Low-temperature electrical resistivity}\label{resistivity}

At ambient pressure, CeRuPO orders ferromagnetically at low temperatures. The large magnetic anisotropy is expected to induce a gap in the magnetic excitation spectrum. Below $T_{C}$, the electrical resistivity $\rho(T)$ of CeRuPO is indeed well described by an equation considering the residual resistivity $\rho_{0}$, the electron-electron scattering leading to the well-known $AT^2$ term, and an exponential term corresponding to the contribution of the magnon modes [see Eq.\,(\ref{eqn:rho with FM gap})]. A fit to the temperature-dependent resistivity at ambient pressure describes the data below $0.8T_{C}$ and yields $\rho_{0}= 6.08\,\mu\Omega$cm, an electron-electron scattering cross section $A=0.14\,\mu\Omega$cm/K$^{2}$, and an energy gap of $\Delta=32.8$\,K. The values of the temperature coefficient $A$ and of the residual resistivity $\rho_{0}$ are consistent with those previously published considering only a LFL behavior at very low temperatures.\cite{krellner08} The contribution of the electron-phonon scattering is negligible in the temperature range $T< 0.8\,T_{C}$ since $\theta_{D}\approx 200$\,K and $T_{C} =14$\,K.

\begin{figure}[t!]
    \centering
    \includegraphics[width=0.95\linewidth]{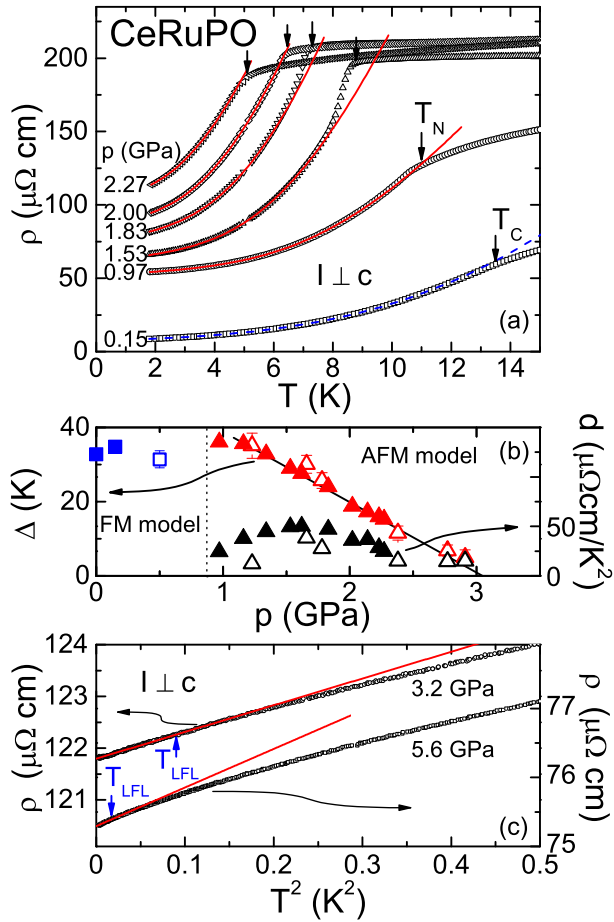}
       \renewcommand{\baselinestretch}{1.2}
    \caption{(a) Temperature dependence of the electrical resistivity of CeRuPO at various pressures. The dashed and solid lines are the fits to the data according to FM and AFM models corresponding to Eqs.\,(\ref{eqn:rho with FM gap}) and (\ref{eqn:rho with AFM gap}), respectively. Pressure dependencies of the parameters extracted from these fits, the magnon excitation gap $\Delta$ [(b), left axis] and for the AFM regime of the phase diagram the parameter $d$ [(b), right axis]. The closed and open symbols indicate the results obtained by employing a piston-cylinder-type and a Bridgman-type pressure cell, respectively. (c) Temperature dependence of the electrical resistivity, $\rho(T^2)$, of CeRuPO in the PM regime at two selected pressures as indicated. The solid lines are fits to the data according to the LFL theory. The arrows mark $T_{\rm LFL}$, the upper temperature limit of the $T^{2}$ dependence of the resistivity. }
    \label{rho(p)_fit with magnon gap}
\end{figure}

Analyzing the specific-heat data of Krellner \textit{et al.} \cite{krellner07} in a similar way using Eq.\,(\ref{eqn:sp heat with FM gap}) and assuming the existence of a FM gap in the magnon excitation spectrum, we obtain a Sommerfeld coefficient $\gamma=78$\,mJ/molK$^{2}$, which is in good agreement with the one reported in Ref.\,\onlinecite{krellner07}. The gap in the FM spin-wave excitation spectrum is estimated to $\Delta=30.2$\,K, which is similar to the value obtained from the analysis of the electrical resistivity ($\Delta=32.8$\,K).

\begin{figure}[t!]
    \centering
    \includegraphics [width=0.85\linewidth]{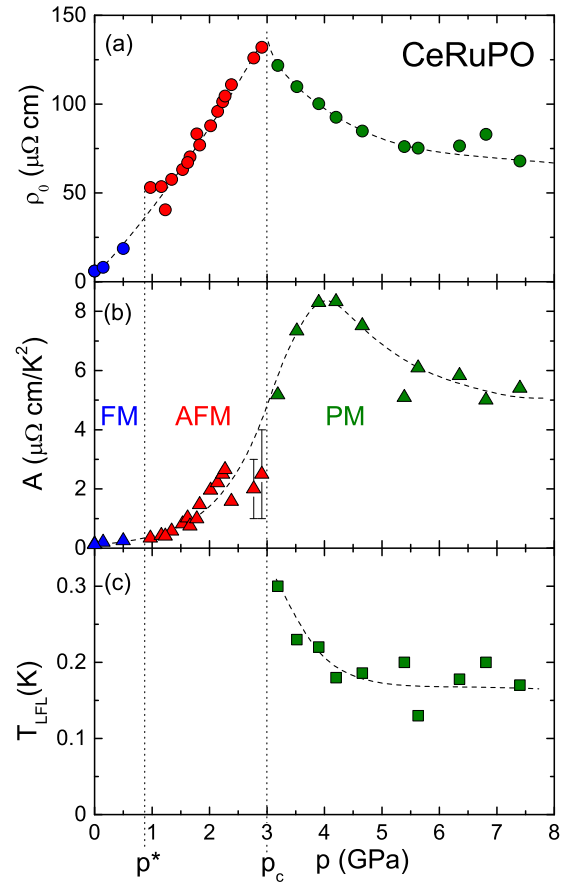}
    \renewcommand{\baselinestretch}{1.2}
    \caption{Pressure dependencies of the parameters $\rho_{0}$, $A$, and $T_{\rm LFL}$. The data in the FM and AFM states are obtained from fits of the $\rho(T)$ data by Eqs.\,(\ref{eqn:rho with FM gap}) and (\ref{eqn:rho with AFM gap}), respectively. The results in the PM regime correspond to fits of the low-temperature resistivity by $\rho(T)=\rho_{0}+A T^{2}$. The vertical dotted lines mark the estimated pressure for the change from the FM to the AFM ground state and the critical pressure where the magnetic order is suppressed to zero temperature $p^*$ and $p_c$, respectively. The dashed lines are guides to the eye.}
    \label{low temp analysis(p)}
\end{figure}

Motivated by the consistent description of the resistivity and specific heat suggesting the formation of a gap in the magnetic excitation spectrum, we analyze the behavior of the low-temperature resistivity under pressure by considering impurity, electron-electron, and electron-magnon scattering. Since the magnon spectrum depends on the type of magnetic ordering, the resistivity data in the FM phase ($p< p^{*}$) were fitted using Eq.\,(\ref{eqn:rho with FM gap}) and in the AFM phase ($p>p^{*}$) according to Eq.\,(\ref{eqn:rho with AFM gap}) (see Appendix\,\ref{apx_mag} for further details). The data were fitted up to $0.8T_{C(N)}$. We find a very good description of the temperature dependence of the resistivity in the ferromagnetically as well as in the antiferromagnetically ordered state as shown in Fig.\,\ref{rho(p)_fit with magnon gap}(a). This indicates that the scattering of the conduction electrons off magnetic excitations is an important contribution to the electrical resistivity which cannot be neglected in the magnetically ordered phases. The gap $\Delta$, depicted in Fig.\,\ref{rho(p)_fit with magnon gap}(b), decreases with increasing pressure and extrapolates to zero at the critical pressure $p_{c} \simeq 3$\,GPa. For further details we refer to Appendix\,\ref{spinwave}.

In the PM regime ($p>p_{c}$), the low-temperature electrical resistivity of CeRuPO follows $\rho(T)=\rho_{0}+A T^{2}$, indicating LFL behavior. Figure\,\ref{rho(p)_fit with magnon gap}(c) displays representative data at $3.2$ and $5.6$\,GPa. We denote the upper temperature limit of the $T^{2}$ dependence in the resistivity by $T_{\rm LFL}$. The pressure dependencies of $\rho_{0}$, $A$, and $T_{\rm LFL}$ are depicted in Fig.\,\ref{low temp analysis(p)} for the whole investigated pressure range. The vertical dotted lines mark $p^*$ and the critical pressure, $p_{c}$, where the magnetic transition is completely suppressed. The residual resistivity increases from $\rho_{0}\approx 6\,\mu\Omega$\,cm at ambient pressure to about $120\,\mu\Omega$\,cm at $p=2.6$\,GPa, i.e., an increase by a factor of $20$. It is noticeable that $\rho_{0}(p)$ exhibits a maximum at $p_{c}= 3$\,GPa. At elevated pressures, $\rho_{0}(p)$ decreases to a value well above the one found at ambient pressure. In the LFL theory, the coefficient of the $T^{2}$ term in the resistivity, $A$, is a measure of the quasiparticle-quasiparticle scattering cross section. Similar to $\rho_{0}(p)$, the $A$ coefficient increases from $0.12$ to $8.3\,\mu\Omega$\,cm/K$^{2}$ between $0$ and $4$\,GPa, where it exhibits a maximum.  Above this pressure, a decrease in $A(p)$ to $3.98\,\mu\Omega$\,cm/K$^{2}$ at $p= 6.8$\,GPa is found. The maximum in $A(p)$ is situated inside the PM region, away from the critical pressure $p_c$, in contrast to the expectation for a QCP. In the lower panel of Fig.\,\ref{low temp analysis(p)}, the evolution of $T_{\rm LFL}$ for pressures above $p_{c}$ is depicted. The temperature range where the resistivity exhibits a $T^{2}$ dependence becomes smaller upon increasing pressure. Above $4$\,GPa, $T_{\rm LFL}$ is almost constant. In particular, we do not find any non-Fermi-liquid behavior in the low-temperature electrical resistivity at the critical pressure. This observation is consistent with the first-order-like disappearance of $T_N$ at $p_c$ and the absence of a pressure-induced QCP in CeRuPO. We note that application of a static magnetic field, $B\parallel c$, at 4.6\,GPa where $A(p)$ is close to its maximum, enlarges the temperature range where LFL behavior is observed in the resistivity (see Appendix\,\ref{parmagnetic} for details).

\section{DISCUSSION}\label{discussion}

The physical properties of CeRuPO are highly sensitive to external pressure. Hydrostatic pressure does not only suppress the magnetic order, but also leads to a change in the type of the magnetic ordering from FM to AFM. At ambient pressure, the Ce $4f$ moments lie along the crystallographic $c$ direction, while the easy magnetization direction is perpendicular to the $c$ axis. This behavior is explained by the competition of the anisotropic RKKY exchange interaction and the single-ion anisotropy due to the CEF interaction, which splits the Ce $5/2$ multiplet into three doublets.\cite{krellner08} External pressure has a strong effect on the Kondo temperature and the CEF-level scheme in CeRuPO.\cite{Macovei09} For the following discussion we will invoke the high-temperature phase diagram and the detailed analysis of the high-temperature electrical resistivity of Macovei \textit{et al.}\cite{Macovei09} in order to correlate them with our low-temperature results.

Generally, in the case when the Kondo temperature $T_K$ is much smaller than the splitting $\delta$, between the ground-state and the first excited doublet the position of the high-temperature maximum ($T_{\rm max}$) in the electrical resistivity can be taken as a measure of the Kondo temperature.\cite{Lassailly85} In Ce-based Kondo lattices $T_K$ and, therefore, $T_{\rm max}$ typically shift to higher temperatures upon application of external pressure. The opposite is observed in CeRuPO: $T_{\rm max}(p)$ initially shifts to lower temperatures upon increasing the pressure.\cite{Macovei09} Such an unexpected negative shift might be explained if we assume the splitting $\Delta$ between the ground-state and the first excited CEF level to be less than $\sim5T_K$ and to decrease with increasing pressure. Then $T_{\rm max}(p)$ depends on both, $T_K$ and $\Delta$, and may shift to lower temperatures with increasing pressure. This suggests that the excited CEF level(s) play an important role for the low-temperature properties of CeRuPO and have to be included in a theoretical scenario to explain the changes observed in the magnetic ground-state properties.\cite{Macovei09} The negative pressure shift of $T_{\rm max}(p)$ is consistent with a reduction of the CEF anisotropy in the low-pressure regime ($p\lesssim1.3$ GPa).\cite{hanzawa85,Macovei09} Above $p\approx 1$\,GPa, the observation of two maxima (shoulders) can be attributed to the Kondo scattering on the ground-state doublet and on the first excited CEF level. We note that this pressure is close to the region where we observe the change from the FM to the AFM order at low temperatures (see the phase diagram in Fig.\,\ref{PhD}). Upon further increasing pressure the two features in the resistivity exhibit distinct pressure dependencies: the high-temperature maximum increases almost linearly with increasing pressure, while the low-temperature feature decreases in temperature up to 2\,GPa.\cite{Macovei09} Above this pressure the low-temperature shoulder shows a temperature independent behavior.\cite{Macovei09} This is also the pressure range, $p\gtrsim2$\,GPa, where we observe a different type of field-polarized FM order (see Sec.\,\ref{AFM_state}). Thus, we take this apparent connection between the high-temperature resistivity features, which indicate the pressure evolution of the corresponding energy scales, and the low-temperature magnetic phases as a hint of a pressure-induced change in the CEF-level scheme and in the size and/or anisotropy of the magnetic exchange interaction. Indeed, our data indicate a reduced magnetic anisotropy with increasing pressure (see Sec.\,\ref{AFM_state}). This is also consistent with structural data, which reveal a decrease of the structural anisotropy.\cite{Hirai13}

At ambient pressure, the CEF ground state of CeRuPO is presumably a $\Gamma_6$ doublet.\cite{krellner08} Based on the pressure evolution of the high-temperature resistivity features, which are directly related to the involved energy scales, one might speculate that one of the two excited $\Gamma_7$ doublets moves down in energy and becomes the new ground state upon increasing pressure. This could give rise to a change in the direction of the ordered moments from pointing along the crystallographic $c$ direction to an in-plane orientation. Furthermore, the anisotropic RKKY exchange interaction between the localized moments, which is responsible for the type of magnetic ordering, is expected to react sensitively on anisotropic changes of the lattice parameters. In CeRuPO, the $c$ axis is more compressible than the $a$ axis, leading to a reduction of the crystalline anisotropy upon application of pressure.\cite{Hirai13} The $c/a$ decreases linearly from 2.051 at ambient pressure to 2.024 at 5.9\,GPa.\cite{Hirai13} Thus, the evolution of the ground state of CeRuPO from FM to AFM and finally PM is most likely based on the complex changes of the CEF-level scheme and the RKKY exchange interaction. In the latter case, the reduced anisotropy might play an important role. In particular, the sensitivity and the different response of the AFM state to the application of a static magnetic field at the different pressures is a direct response to this complex behavior (see Sec.\,\ref{AFM_state}).

While external pressure reduces the anisotropy in CeRuPO,\cite{Hirai13} isoelectronic substitution of Ru by Fe in CeRu$_{1-x}$Fe$_x$PO enhances the anisotropy significantly. Upon increasing the external pressure both, the $a$- and $c$-axis lattice parameters shrink, while upon increasing the Fe concentration in CeRuPO, $a$ decreases and $c$ increases leading to an enhancement of the anisotropy.\cite{Kitagawa12} This increase of the crystalline anisotropy correlates with a decrease of the dimensionality of the FM fluctuations.\cite{Kitagawa13} NMR data indicate dominant three-dimensional FM correlations at ambient pressure which give rise to the FM ordering in CeRuPO.\cite{Kitagawa13} With increasing the Fe concentration, the out-of-plane fluctuations are significantly suppressed and the FM fluctuations become two dimensional near a FM QCP at $x\approx0.86$ in CeRu$_{1-x}$Fe$_x$PO.\cite{Kitagawa13} For the observed behavior, Kitagawa \textit{et al.}\ suggest a new and different tuning mechanism, namely, that $T_C$ is suppressed to a FM QCP by tuning the dimensionality of the magnetic correlations. In contrast to the Fe substitution, by increasing the external pressure we drive CeRuPO to the opposite direction making the material less  anisotropic. Adopting now the scenario proposed by Kitagawa \textit{et al.}, the different response to \textit{pressure} observed in our external-pressure investigation is no surprise and in line with the iron-substitution studies,\cite{Kitagawa12,Kitagawa13} even though both the external pressure and Fe substitution lead to a reduction of the unit-cell volume.\cite{Kitagawa12,Hirai13} Recently, we became aware of an NMR study on CeRuPO under external pressure.\cite{Kitagawa14} The general conclusions of that work are consistent with our results, though, we note that a magnetic field had to be applied to take the NMR data.

\section{SUMMARY}\label{summary}

In summary, CeRuPO is one of the rare examples of a FM Kondo-lattice system. Application of hydrostatic pressure suppresses the magnetic-ordering temperature. The type of magnetic order changes from FM to AFM at about $p^*\approx0.87$\,GPa, where possibly a tricritical point exists. The N\'{e}el temperature disappears abruptly around $p_c\approx 3$\,GPa in a first-order-like fashion. The electrical resistivity exhibits a LFL behavior in the whole investigated pressure range, which is consistent with the absence of a QCP at $p_c$ in CeRuPO. The change in the type of the magnetic ordering seems to be intricately related to the pressure evolution of the CEF-level scheme and of the anisotropy of the magnetic exchange interaction. Even though CeRuPO possesses an AFM ground state between $p^*$ and $p_c$, the application of a small magnetic field ($B\parallel c$) induces a field-polarized state. We find further evidence that the FM correlations at high pressures ($p\gtrsim 2$\,GPa) are different in nature compared with the fluctuations at lower pressure. The FM fluctuations at high pressures might be related to FM in-plane and significant AFM out-of-plane correlations. Our results evidence that CeRuPO is close to ferromagnetism in the entire pressure range $p<p_c$, in particular, also where it possesses an AFM ground state. Furthermore, our data lead us to speculate that not only is the type of magnetic ordering changing, but also the direction of the ordered moments. To confirm this suggestion, further microscopic magnetic measurements on CeRuPO in zero magnetic field under external pressure would be desirable.

\appendix

\section{Scattering contributions to the resistivity}\label{apx_mag}

The interaction between magnons and conduction electrons in a magnetic metal contributes to the electrical resistivity along with the electron-phonon, electron-electron, and electron-impurity scattering. It has been shown that at temperatures below the magnon energy gap $\Delta$, the electron-magnon resistivity falls off exponentially with decreasing temperature \cite{mackintosh63,andersen smith79}. It was suggested that the temperature dependence of the resistivity in a FM metallic system with a magnon energy gap is proportional to $T^{2}e^{-\Delta/T}$ \cite{mackintosh63}. Andersen and Smith \cite{andersen smith79} predicted that, at low temperatures, the resistivity can be described by $\Delta \rho(T) \propto Te^{-\Delta/T}$. In our analysis, we follow the approach of Ref.\,\onlinecite{andersen smith79}.

In general, the resistivity associated with electrons scattering off an arbitrary type of boson excitation, e.g., magnon or phonon, is given by \cite{andersen smith79}

\begin{equation}
    \label{eqn:rho with gap after Andersen}
    \rho_{B}(T)=\cfrac{m\pi N(0)}{ne^{2}}\int_{0}^{2k_{F}}\!\cfrac{k^{3}}{k^{2}_{F}}\,dk \int \cfrac{d\Omega _{\vec{k}}}{4\pi}| g_{\vec{k}}|^{2}\cfrac{\cfrac{\hbar\omega _{\vec{k}}}{k_{B}T}}{4\sinh^{2}\cfrac{\hbar\omega _{\vec{k}}}{2k_{B}T}},
\end{equation}\vspace{0.3cm}

\noindent where $n=k_{F}^{3}/3\pi^{2}$ is the number density of the charge carriers, $N(0)=(mk_{F})/(2\pi^{2}\hbar^{2})$ is the density of states per spin at the
Fermi level, $2k_{F}$ represents the maximum wave-vector transfer, $g_{\vec{k}}$ is the electron-boson coupling, and 	 $\hbar\omega _{\vec{k}}$ is the boson energy for a given wave vector $\vec{k}$. The application of Eq.~(\ref{eqn:rho with gap after Andersen}) leads, for electron-phonon scattering, to the Bloch law $\Delta \rho(T)\propto T^{5}$ since $| g_{\vec{k}}|^{2} \propto k$ \cite{andersen smith79}. In the electron-magnon case, when the gap $\Delta=0$, the resistivity is proportional to $T^{2}$ and $T^{5}$ for an isotropic ferromagnet and antiferromagnet, respectively \cite{andersen smith79}. If a gap is present in the magnon spectrum, in the case of an anisotropic FM material, the electron-magnon coupling is independent of ${\vec{k}}$ and the energy dispersion relation of the magnons is expressed by $\hbar \omega _{\vec{k}}=\Delta + D k^{2}$, where $D$ is the spin-wave stiffness \cite{andersen smith79}. The total resistivity in the ordered state in this case is expressed as \cite{andersen smith79}

\begin{equation}
\label{eqn:rho with FM gap}
\rho(T)=\rho_{0}+AT^{2}+ bT\Delta\left(1+\frac{2T}{\Delta}\right)e^{-\Delta/T},
\end{equation}\vspace{0.2cm}

\noindent where $\rho_{0}$ is the residual resistivity, the second term accounts for the electron-electron scattering, and the exponential term corresponds to the contribution of the magnon modes. $b$ is a constant for a given material depending on the spin-wave stiffness and the strength of the electron-magnon coupling.

Accordingly, the magnetic contribution to the specific heat of CeRuPO for $T<T_{C}$ can be written as

\begin{equation}
\label{eqn:sp heat with FM gap}
C^{4f} (T)= \gamma T + C_{gap}(T),
\end{equation}\vspace{0.2cm}

\noindent where $\gamma T$ is the usual electronic term and

\begin{equation}
\label{eqn:sp heat gap}
C_{gap}(T)= \beta(\Delta^{2}/\sqrt{T}+3\Delta\sqrt{T}+5T^{3/2})e^{-\Delta/T}
\end{equation}

\noindent describes the contribution due to a FM spin-wave excitation spectrum with an energy gap $\Delta$ \cite{coqblin77}. Here, $\beta$ is a constant similar to $b$. The $4f$ contribution to the specific heat is described for $T\leq 0.8\,T_{C}$ by Eq.~(\ref{eqn:sp heat with FM gap}).

In the case of an anisotropic AFM system, the electron-magnon coupling depends on ${\vec{k}}$ and the energy dispersion relation is given by $\hbar \omega _{\vec{k}}=\sqrt{\Delta^{2}+D k^{2}}$. Thus, the resistivity for an AFM system for $T\ll \Delta$ is given by \cite[e.g.,][]{Continentino01}

\begin{equation}
\begin{split}
\label{eqn:rho with AFM gap}
\rho(T)=\rho_{0}+AT^{2}+ d\Delta^{3/2}T^{1/2}e^{-\Delta/T} \times\\ \times \left[1+\frac{2}{3}\left(\frac{T}{\Delta}\right)+\frac{2}{15}\left(\frac{T}{\Delta}\right)^{2}\right],
\end{split}
\end{equation}\vspace{0.2cm}

\noindent where the exponential term corresponds to the contribution of AFM magnons with a gap, $\Delta$, in the spin-wave excitation spectrum and the coefficient, $d$, is related to the spin-wave stiffness and the strength of the electron-magnon coupling.

\section{Spin-wave gap and stiffness in the antiferromagnetic state} \label{spinwave}

The size of the gap, $\Delta$, obtained from the fits according to Eq.\,(\ref{eqn:rho with AFM gap}) in the AFM regime is plotted as a function of pressure in Fig.\,\ref{rho(p)_fit with magnon gap}(b). There is a clear correlation between $T_{N}(p)$ and $\Delta(p)$ in the AFM state, i.e., both decrease with increasing pressure. In particular, for $p>p^{*}$, the gap decreases linearly with increasing pressure. An extrapolation of $\Delta (p)\rightarrow 0$ yields the critical pressure $p_{c} \simeq 3$\,GPa. This value coincides with the critical pressure where $T_{N}$ is completely suppressed. Nevertheless, the values of $\Delta$ obtained at $p=2.7$ and $2.9$\,GPa deviate slightly from the linear-in-$p$ dependence of $\Delta$. Therefore, the spin-wave gap might also close suddenly at $p_{c}$. The parameter, $d(p)$, which is proportional to the strength of the electron-magnon coupling and to the inverse of the spin-wave stiffness [see Fig.\,\ref{rho(p)_fit with magnon gap}(b) and Eq.\,(\ref{eqn:rho with AFM gap})] remains finite and does not change significantly upon approaching the critical pressure. Since the strength of the electron-magnon coupling is expected to increase with pressure due to an increasing hybridization between the 4$f$ and the conduction electrons, the nearly pressure-independent $d(p)$ might hint at an increase of the spin-wave stiffness with pressure for $p>p^{*}$. However, with only resistivity data, it is not possible to discern between the effects of these two parameters.

\section{Effect of a magnetic field in the paramagnetic state}
\label{parmagnetic}

\begin{figure}[t!]
    \centering
    \includegraphics [width=0.85\linewidth]{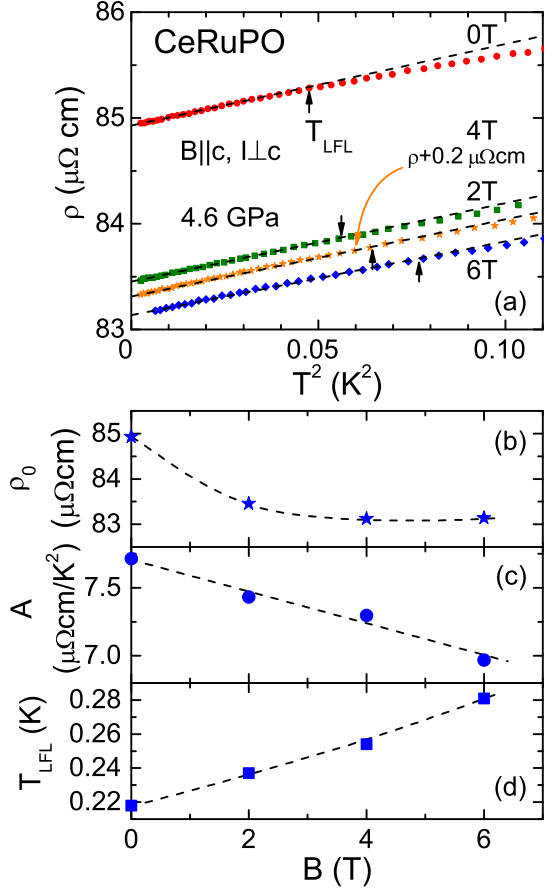}
    \renewcommand{\baselinestretch}{1.2}
    \caption{(a) Low-temperature electrical resistivity of CeRuPO at $p =4.6\,\rm GPa$ plotted against $T^{2}$ for different magnetic fields. For clarity, the data taken in 4\,T have been shifted by 0.2\,$\mu\Omega {\rm cm}$. The dashed lines describe $\Delta\rho(T)\propto T^{2}$. The magnetic field dependence of the parameters (b) $\rho_{0}$ and (c) $A$ obtained from fitting $\rho(T)$ with $\rho(T)=\rho_{0}+A T^{2}$. (d) Evolution of the upper temperature limit, $T_{\rm LFL}$, of the $T^{2}$ dependence of $\rho(T)$ with magnetic field. The dashed lines are guides to the eye.}
    \label{rho_4.6GPa}
\end{figure}

Figure\,\ref{rho_4.6GPa}(a) presents the temperature dependence of $\rho$ at $p = 4.6$\,GPa, as $\rho(T)$ {\it vs.} $T^{2}$ for various static magnetic fields up to $B=6$\,T ($B\parallel c$). The linear dependence of $\rho(T)$ {\it vs.} $T^{2}$ shows that at low temperatures the electrical resistivity has a quadratic temperature dependence for the entire investigated magnetic-field range. This  behavior is a hallmark of a LFL ground state. The deviation of $\rho(T)$ {\it vs.} $T^{2}$ from the straight line denotes the crossover temperature $T_{\rm LFL}$. The results of an analysis of the $\rho(T)$ data according to $\rho(T)=\rho_{0}+A T^{2}$ for various static magnetic fields are summarized in Figs.\,\ref{rho_4.6GPa}(b)--\ref{rho_4.6GPa}(d). As shown in Fig.\,\ref{rho_4.6GPa}(b), $\rho_{0}(B)$ decreases slightly for fields up to $2$\,T, followed by an almost field independent behavior. The temperature coefficient $A$ decreases linearly upon increasing the magnetic field [Fig.\,\ref{rho_4.6GPa}(c)], by about 10\% at 6\,T, suggesting a decrease of the heavy quasiparticle mass $m^{*}$ with increasing $B$. In Fig.\,\ref{rho_4.6GPa}d, the evolution of $T_{\rm LFL}$ with magnetic field is depicted. As the field increases, the temperature range of the $T^{2}$ dependence of $\rho(T)$ expands, indicating a stabilization of the LFL state with increasing the magnetic field.

\end{document}